\documentclass[reprint,amsmath,amssymb,aps,]{revtex4-2}
\usepackage{graphicx}
\usepackage{dcolumn}
\usepackage{bm}
\usepackage{lipsum}
\usepackage[colorlinks=true,breaklinks=true,linkcolor=blue,citecolor=blue,urlcolor=blue]{hyperref}
\usepackage{xcolor}

\begin{document}

\preprint{APS/123-QED}

\title{Contribution of fictitious forces to polarization drag in rotating media}

\author{Julien Langlois}
\email{julien.langlois@laplace.univ-tlse.fr}
\author{Renaud Gueroult}
\affiliation{LAPLACE, Université de Toulouse, CNRS, INPT, UPS, 31062 Toulouse, France}

\date{\today}

\begin{abstract}
    Models for polarization drag - mechanical analog of the Faraday effect - are extended to include inertial corrections to the dielectrics properties of the rotating medium in its rest-frame. Instead of the Coriolis-Faraday term originally proposed by Baranova \& Zel'dovich, inertia corrections due to the fictitious Coriolis and centrifugal forces are here derived by considering the effect of rotation on both the Lorentz and plasma dielectric models. These modified rest-frame properties are subsequently used to deduce laboratory properties. Although elegant and insightful, it is shown that the Coriolis-Faraday correction inferred from Larmor's theorem is limited in that it can only capture inertial corrections to polarization drag when the equivalent Faraday rotation is defined at the wave frequency of interest. This is notably not the case for low frequency polarization drag in a rotating magnetized plasma, although it is verified here using the more general phenomenological models that the impact of fictitious forces is in general negligible in these conditions. 
\end{abstract}

\maketitle

\section{\label{sec:introduction}Introduction} 

In a medium supporting right ($\textsf{R}$) and left ($\textsf{L}$) circularly polarized eigenmodes, a linearly polarized wave sees its polarization rotate when the refractive indices $n_\textsf{R}$ and $n_\textsf{L}$ of these modes differ~\cite{fowles1989,barron2009}. Indeed, the difference in indexes introduces a phase shift between the two eigenmodes such that the change in polarization angle per unit length of a linearly polarized wave obtained from the combination of these modes is
\begin{equation} \label{eq:polarization_rotation_angle}
    \delta(\omega) = \frac{\omega}{2c} \Big[n_\textsf{L}(\omega) - n_\textsf{R}(\omega) \Big].
\end{equation}
This phenomenon is referred to as circular birefringence, a manifestation of optical activity, and $\delta$ is the specific rotatory power~\cite{fowles1989}. Circular birefringence was most notably demonstrated by Faraday in 1845~\cite{faraday1846} for light propagating along the direction of an externally imposed magnetic field, and polarization rotation due to a background magnetic field is commonly referred to as Faraday rotation. 

Thomson postulated in 1885 that an equivalent phenomenon should take place in rotating media~\cite{thomson1885}. The idea was revisited a few decades latter by Fermi~\cite{fermi1923}, who then drew an analogy with the Fresnel drag experienced by light propagating through a dielectrics in linear motion~\cite{fresnel1818,jones1975,player1975}. Fermi accordingly coined the name \emph{polarization drag} for this mechanically-induced polarization rotation. The existence of this effect was only successfully demonstrated experimentally by Jones in 1976 by studying a laser beam propagating through a rotating glass rod~\cite{jones1976}. Beyond demonstrating unequivocally the effect, Jones's experimental findings were found to support the corrections made at the time by Player to Fermi's theory to account for dispersion~\cite{player1976}. In light of this apparent agreement, Player's theory has since then been used to examine the effect of rotation on waves carrying orbital angular momentum~\cite{gotte2007}. It was also used more recently to model the effect of rotation on media exhibiting an anisotropic dielectric response in their proper reference frames~\cite{gueroult2019}, with the goal of studying polarization drag in rotating magnetized plasmas~\cite{gueroult2019,gueroult2020,gueroult2023}. These new developments were key in underlining both the possible effect of polarization drag in pulsar magnetospheres and how it may enable determining the rotation direction of pulsars~\cite{gueroult2019}, and the opportunities rotating magnetized plasmas may offer for the development of high performance non-reciprocal elements~\cite{gueroult2020}.

In building on Player's work~\cite{player1976}, the above theoretical contributions on polarization drag, however, all have in common that they assume that the dielectric properties in the medium’s proper reference frame (or "rest frame") are not modified by rotation. The fact that this simplification may be a problem was first raised by Baranova \& Zel'dovich~\cite{baranova1979}, in relation with Jones' and Player's work~\footnote{To be exact the issue of a possible effect of inertia on the constitutive relations appears to have first been raised by M. H. L. Pryce in a private communication to R. V. Jones in October 1975~\cite{JonesArchives}}. On the basis of Larmor's theorem asserting a first order equivalence between the effect of a uniform rotation at angular frequency $\boldsymbol{\Omega}$ and of a homogeneous magnetic field $\mathbf B^* = 2m\boldsymbol{\Omega}/q$ on the dynamics of a particle of charge $q$ and mass $m$~\cite{larmor1900,brillouin1945}, they proposed to add an extra term  - the Coriolis-Faraday term - to Player's formula. Leaning on Larmor's theorem, they postulated that this additional term should take the form of an equivalent Faraday rotation due to the magnetic field $\mathbf B^*$. However, while the estimate Baranova \& Zel'dovich obtained for Jones' experiment appear to question the classic $n_g -n_{\phi}$ Fresnel hybrid scaling derived by Player~\cite{player1976} and identified by Jones~\cite{jones1976}, this apparent contradiction, and from there uncertainty both on the effect of inertia on polarization drag and on Player's formula, remains. This was notably noted by Woerdman~\emph{et al.} \cite{woerdman1992} in studying analogies between Faraday rotation and polarization drag~\cite{nienhuis1992}.  

To complicate things further, as pointed by Baranova \& Zel'dovich themselves~\cite{baranova1979}, the proposed Coriolis-Faraday analogy for polarization drag only holds when the ionic response is negligible. Because it is only concerned with the Coriolis force, it also only holds at frequencies where centrifugal effects are negligible. For these reasons it seems inapplicable to capture inertia effects on polarization drag for the case of low frequency waves in a rotating magnetized plasma that was recently considered~\cite{gueroult2019,gueroult2020}. If for no other reason, the direct application of this analogy for a two species plasma would also lead to two different $\mathbf B^*$ for the same $\boldsymbol\Omega$. Yet, this inapplicability of the Faraday-Coriolis analogy does not \emph{a priori} guarantee that inertial effects are not important. Here we address this issue by considering the effect of "fictitious" forces on the microscopic description of the medium - similarly to the work of Shiozawa~\cite{shiozawa1973,shiozawa1974} - to obtain the medium's dielectric response corrected for inertia in its rest-frame. These inertia-corrected rest-frame properties can then be used to derive general expression for polarization drag. Note that while they have in common to compute the dielectric properties in the rotating frame to deduce polarization drag properties in the laboratory frame, this work differs from the study of polarization drag in super-rotors~\cite{Steinitz2020,Milner2021,Tutunnikov2022} in that rotation is here considered around a common axis, whereas each super-rotors molecule rotates around its own axis.

This paper is organized as follows. First we recall in Section~\ref{sec:theory} how Player's method~\cite{player1976} allows determining the polarization rotation observed in the laboratory frame from rest-frame dielectric properties, and how it has since then been extended to media with gyrotropic rest-frame properties. Then, in Section~\ref{sec:models}, phenomenological models for Lorentz dielectrics and plasmas are used to determine the dielectric response of these media when rotating at angular frequency $\boldsymbol\Omega$ and in the presence of a background magnetic field $\mathbf{B}_0$. It is shown that this response conveniently takes a form analogous to that of a gyrotropic medium if $\boldsymbol\Omega||\mathbf{B}_0$, allowing to use Player's extended method to derive laboratory-frame properties. Putting these pieces together in Section~\ref{sec:drag}, the limits of applicability of the Coriolis-Faraday analogy are first confirmed. The new models are then used to quantify inertial corrections to low-frequency polarization drag in a rotating magnetized plasma, where Coriolis-Faraday is inapplicable. The main findings of this study are finally summarized in Section~\ref{sec:conclusions}. 

\section{\label{sec:theory}Circular birefringence \\ from rest-frame susceptibility}

In this section we briefly recall how circular birefringence induced by rotation - \emph{i.e.} polarization drag - can be derived from the rest-frame dielectrics properties of the medium, and underline under which hypotheses on the form of the rest-frame susceptibility tensor and wave vector analytical expressions are readily available. 

For this we consider a medium that rotates with constant angular frequency $\boldsymbol\Omega$. We note respectively $\Sigma$ and $\Sigma'$ the laboratory frame and the medium rest-frame, and use a prime notation to refer to variables expressed in $\Sigma'$. Without loss of generality, we take $\hat{\mathbf z} = \boldsymbol\Omega/|\boldsymbol\Omega|$ the unit vector along the axis common to both $\Sigma'$ and $\Sigma$. 

Propagation properties observed in $\Sigma$ can be derived from the dielectric properties in $\Sigma'$ following the method first introduced by Player~\cite{player1976}. This method uses as inputs the constitutive relations in $\Sigma'$. Then, by considering the Lorentz transform of the fields ($\mathbf{D}'$, $\mathbf{B}'$, $\mathbf{E}'$, $\mathbf{H}'$) in $\Sigma'$ to ($\mathbf{D}$, $\mathbf{B}$, $\mathbf{E}$, $\mathbf{H}$) in $\Sigma$ for the instantaneous velocity ${\mathbf v = \boldsymbol\Omega \times \mathbf r}$~\cite{minkowski1908}, one obtains equivalent constitutive relations in $\Sigma$. Finally, Maxwell's equations in the laboratory frame can be used to derive a wave equation ${\underline{\boldsymbol{\mathcal D}}(\underline{\boldsymbol \chi}', \omega, \mathbf k)\cdot \mathbf F = \mathbf 0}$ for a given field $\mathbf F$. The non-trivial solutions ${\text{det}\left[\underline{\boldsymbol{\mathcal D}}(\underline{\boldsymbol \chi}', \omega, \mathbf k)\right] = 0}$ are the eigenmodes as seen from the laboratory frame, with corresponding dispersion relations.

A number of restrictive hypotheses are however needed to make this derivation tractable. Player's original derivation assumed for instance an isotropic dielectric with rest-frame susceptibility ${\chi'_\parallel(\omega')=\chi_\parallel(\omega')}$, and a wave vector $\mathbf{k}=k\hat{\mathbf z}$ aligned with the rotation axis. Under these assumptions he showed that the eigenmodes in the laboratory frame are circularly polarized with waves indexes
\begin{equation} \label{eq:birefringence_iso}
	n_\textsf{R/L}^2(\omega) = 1 + \chi_\parallel(\omega’) \mp \frac{2\Omega}\omega\chi_\parallel(\omega’)
\end{equation}
with ${\omega' = \omega \mp \Omega}$ the Doppler shifted wave frequency~\cite{mashhoon1989}. In the limit of slow rotation $\Omega\ll\omega$ and for $n_{\phi}(\omega')\neq0$ the wave index difference then leads to first order to a polarization rotation 
\begin{equation}
\label{eq:theta_rig}
    \delta_\textsf{rig}(\omega) = {\frac{\Omega}c\left[n_{g}(\omega) - \frac{1}{n_{\phi}(\omega)} \right ]}
\end{equation}
where ${n_{\phi} = \sqrt{1+\chi_\parallel}}$ and ${n_{g} = n_\phi + \omega (dn_\phi/d\omega)}$ are respectively the phase and group index. The index $\textsf{rig}$ stands here for "rigid body" polarization rotation, in the sense that inertial effects are neglected. Indeed $\chi_\parallel$ is here simply the susceptibility in the medium at rest, since as indicated earlier Player considered the dielectric properties to be unaffected by rotation. Examining Eq.~\eqref{eq:theta_rig}, two distinct contributions can be identified. One is the kinematic term ${(n_\phi-n_\phi^{-1})}$ initially found by Fermi~\cite{fermi1923}. The other is the dispersive term ${(\omega dn_\phi/d\omega)}$ derived by Player~\cite{player1976}. This dispersive term was later interpreted by Woerdman \textit{et al.} as the intrinsic rotation contribution of the individual microscopic systems~\cite{woerdman1992}.

A number of Player's original assumptions have nevertheless been lifted since then. G{\"o}tte \emph{et al.} considered for instance the case where the wave vector is not aligned with $\bm{\Omega}$~\cite{gotte2007}, so as to underlined the phenomenon of image rotation for orbital angular momentum carrying waves~\cite{Franke-Arnold2011}. More relevant to our work here, Gueroult~\textit{et~al.}~\cite{gueroult2019} generalized Player's derivation to media with an anisotropic Hermitian dielectric tensor in the rest-frame of the form
\begin{equation}
    \underline{\boldsymbol\chi}' = 
    \begin{pmatrix}
        \chi_\perp' & -i\chi_\times' & 0\\ 
        i\chi_\times' & \chi_\perp' & 0\\ 
        0 & 0 & \chi_{\parallel}'
    \end{pmatrix}.
    \label{eq:tensor_form}
\end{equation}
In this case it was  shown that the refractive indices of \textsf{R}- and \textsf{L}-circularly polarized eigenmodes propagating along the rotation axis write
\begin{align} \label{eq:birefringence}
	n_\textsf{R/L}^2(\omega) = 1 &+ \chi_\perp’(\omega’) \pm \chi_\times’(\omega’) \nonumber \\
	&- \frac{\Omega}\omega\Big[\chi_\times’(\omega’)\pm \chi_\perp’(\omega’)\pm \chi_{\parallel}'(\omega’)\Big]
\end{align}
with again ${\omega' = \omega \mp \Omega}$ the Doppler shifted wave frequency. Analytical expression from polarization drag can thus be derived from Eqs.~\eqref{eq:polarization_rotation_angle} and (\ref{eq:birefringence}) for as long as the rest-frame susceptibility tensor takes the form given in Eq.~(\ref{eq:tensor_form}). Although this was done for the purpose of studying rotating magnetized plasmas, we will show in the next section that this same form Eq.~\eqref{eq:tensor_form} fortuitously captures inertia effects in the medium's rest-frame. 

\section{\label{sec:models}Rest frame dielectric response \\ corrected for inertia}

In this section we examine the response of a rotating dielectric to a harmonic wave perturbation in its rest-frame, for two different dielectric models. To do so we write the 
momentum conservation equation in the rest-frame of a rotating dielectric, which by virtue of working in the rotating frame includes fictitious forces. This equation is then linearized to write the current $\mathbf j'$ as a function of the electric field $\mathbf E'$. The rest-frame susceptibility tensor $\underline{\boldsymbol\chi}'$  is then immediately deduced from the definition 
\begin{equation}
    \mathbf j' = \epsilon_0 \underline{\boldsymbol\chi}' \cdot \partial_t\mathbf E'
    \label{eq:tensor_def}
\end{equation}
for a linear dielectric (see \emph{e.g.} Landau \textit{et al.}~\cite{landau2013}).

Note that the mass, density and external magnetic field used in the momentum equation written in the medium rest-frame in this section should be those measured by an observer in $\Sigma'$, and thus should be primed. Similarly, the plasma and cyclotron frequencies used in the rest-frame dielectric tensor should be primed. Yet, to keep things simple, and because our interest is in the non-relativistic limit $\gamma \rightarrow1$, we drop the prime notation, which accounts for an error of at most a factor $\gamma= [1-(v/c)^2]^{-1/2}$.

\subsection{\label{sub:lorentz}Rotating Lorentz oscillator}

\begin{figure}
    \includegraphics[width=0.7\linewidth]{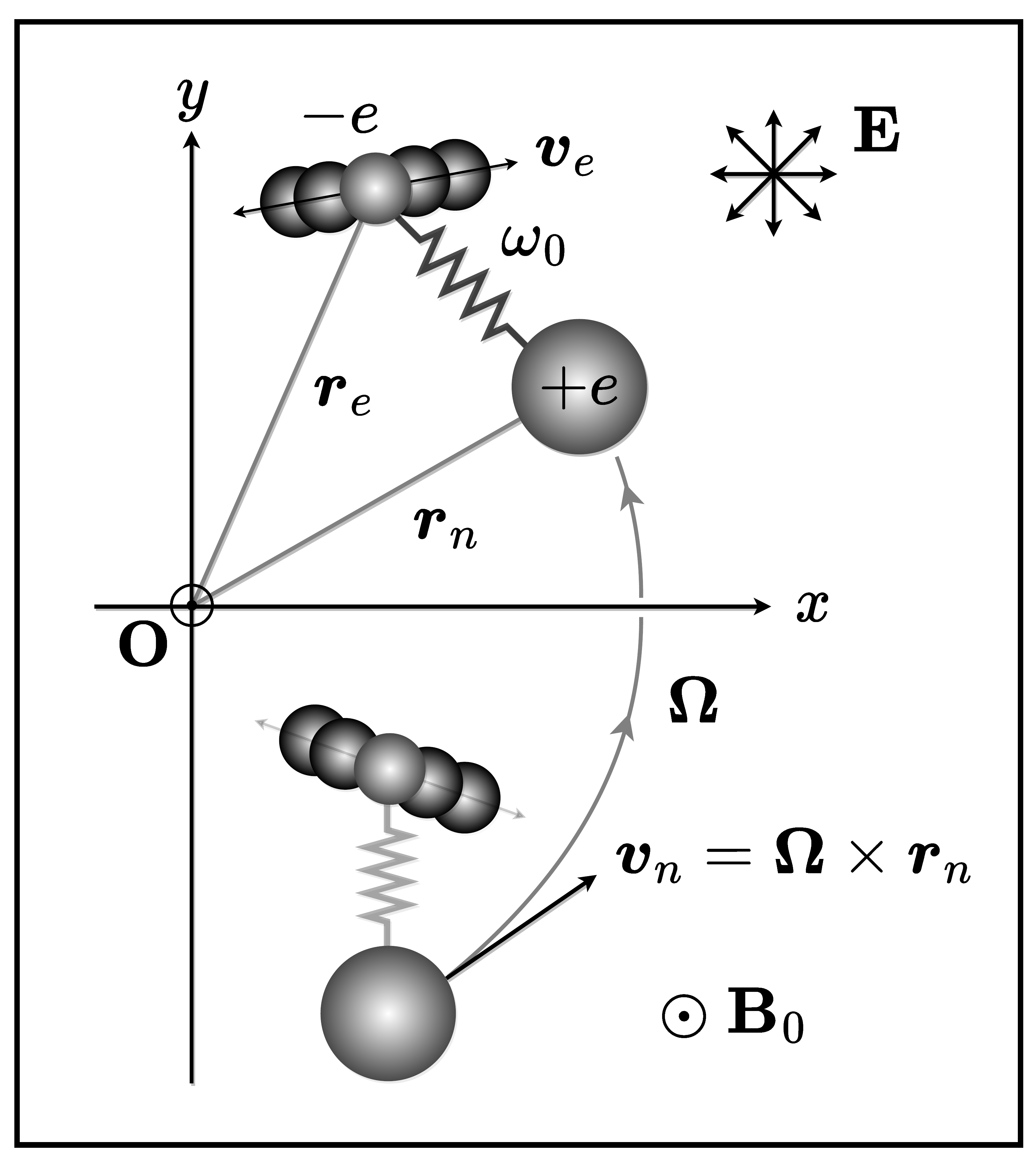}
    \caption{\label{fig:DLS} The rotating Lorentz model in the lab frame.}
\end{figure}

A dielectric medium (solid or gas) can be seen as a cloud of electrons bound to and surrounding a positive nucleus (subscripts $e$ and $n$ respectively), which are deformed in the presence of electric fields. Following Lorentz~\cite{lorentz1916} and Sommerfeld~\cite{sommerfeld1964}, each electron-nucleus couple can be seen as a driven harmonic oscillator with an elastic restoring force. This force is characterized by a frequency $\omega_0$ which represents the proper frequency of the electron moving in the spherically symmetric potential of its nucleus. The macroscopic response is obtained from superposition on all oscillators~\cite{levi2016}. Although it describes a classical, phenomenological behaviour, we note that this Lorentz model is able to capture the essential physics and match the quantum results for the polarization phenomena we are interested in~\cite{berman2010}.

Let ($\boldsymbol r_e', \boldsymbol v_e'$) be the Lagrangian position and velocity of an electron in response to an electromagnetic perturbation $\mathbf E'$ and $\mathbf B'$ in the rest frame $\Sigma'$. Let us also assume that the nuclei are not affected by the perturbations, and are therefore in uniform rotation. One then has $\boldsymbol{v}_n'=\mathbf 0$ and $\boldsymbol{v}_n=\boldsymbol{\Omega}\times\boldsymbol{r}_n$, as shown in Fig.~\ref{fig:DLS}. Taking into account the restoring force, an external background magnetic field ${\mathbf B_0'\approx\mathbf B_0}$ and the fictitious forces due to the rotation, the momentum conservation equation for electrons writes in $\Sigma'$ as
\begin{equation}
    m_e\frac{d\boldsymbol v'_e}{dt}  = -e\left[\mathbf E'+\boldsymbol v'_e\times \tilde{\mathbf{B}}' \right ] - m_e\omega_0^2(\boldsymbol r'_e - \boldsymbol r'_n) + \mathbf{F'}
    \label{eq:momentum}
\end{equation}
where $m_e$ and $-e$ are respectively the mass and charge of electrons, $\tilde{\mathbf{B}}'=\mathbf B'+\mathbf B_0$ is the total magnetic field seeing from rest-frame and
\begin{equation}
    \mathbf F' = -m_e\Big[2\boldsymbol\Omega\times\boldsymbol v'_e + \boldsymbol\Omega\times(\boldsymbol\Omega\times\boldsymbol r'_e) \Big]
\end{equation}
represents the fictitious force exerted on the electrons. Here ${2\boldsymbol\Omega\times\boldsymbol v'_e}$ is the Coriolis force and ${\boldsymbol\Omega\times(\boldsymbol\Omega\times\boldsymbol r'_e)}$ is the centrifugal force, which are respectively linear and quadratic in $\Omega$. 

As we consider only the linear response, the second-order perturbative term $\boldsymbol v_e' \times \mathbf B'$ in Eq.~\eqref{eq:momentum} can be neglected. Considering harmonic time-dependence fields ($\propto e^{-i\omega't}$) and using dyadic notations, the momentum equation can then after some algebra (see Appendix~\ref{app:dyadic}) be rewritten in compact form  
\begin{equation} 
    \frac{m_e}{e}\Big[(\omega'^2+\Omega^2-\omega_0^2){\mathbf I} + i\omega'\boldsymbol\Omega^*\times{\mathbf I}-\boldsymbol\Omega\boldsymbol\Omega\Big]\cdot\boldsymbol v_{e}'=-i\omega'\mathbf E'.
    \label{eq:v_E_dielectric}
\end{equation}
Here we introduced ${\boldsymbol\Omega^* = 2\boldsymbol\Omega-e\mathbf B_0/m_e}$ the effective cyclotron frequency, which takes into account both the medium angular rotation frequency $\Omega$ and the unsigned electronic cyclotron frequency ${\Omega_{ce} = e|\mathbf B_0|/m_e}$. Inverting the tensor on the left side of Eq.~(\ref{eq:v_E_dielectric}) gives $\boldsymbol v_{e}'$ as a function of $\mathbf E'$. 

Now, because the global response of the dielectric is simply obtained by superposition of the responses of each oscillator, one finds in particular ${\mathbf j' = -eN_e \langle\boldsymbol v_e'\rangle}$ with $N_e$ the mean electronic density. Putting these pieces together we get a relation of the form of Eq.~\eqref{eq:tensor_def}, from which the components of the dielectric susceptibility tensor can be readily identified. As detailed in Appendix~\ref{app:inversion}, one finds from Eq.~(\ref{eq:v_E_dielectric})
\begin{subequations} \label{eq:general_dielectric_tensor}
\begin{equation} 
    \underline{\boldsymbol\chi}'= \underline{\boldsymbol\chi}'_\dagger-\frac{\underline{\boldsymbol\chi}'_\dagger\cdot \boldsymbol\varpi\boldsymbol\varpi \cdot \underline{\boldsymbol\chi}'_\dagger}{1+{}^\textsf{T}\boldsymbol\varpi\cdot\underline{\boldsymbol\chi}'_\dagger\cdot\boldsymbol\varpi}
\end{equation}
where 
\begin{equation}
    \underline{\boldsymbol\chi}'_\dagger = -\omega_{pe}^2\frac{\omega'^2\underline{\mathbf I} - \boldsymbol\Omega^*\boldsymbol\Omega^*-i\omega' \boldsymbol\Omega^*\times\underline{\mathbf I}}{\omega'^2-{\Omega^*}^2}
\end{equation}
and $\boldsymbol\varpi$ is a row vector such that
\begin{equation}
    \boldsymbol\varpi\boldsymbol\varpi = -\frac{1}{\omega_{pe}^2}\Big[(\Omega^2-\omega_0^2)\underline{\mathbf I} - \boldsymbol\Omega\boldsymbol\Omega\Big],
\end{equation}
\end{subequations}
with ${\omega_{pe} = [e^2N_e/(\epsilon_0 m_e)]^{1/2}}$ the electron plasma frequency. Although all 9 components of $\underline{\boldsymbol\chi}'$ are in general non-zero, one verifies that $\underline{\boldsymbol\chi}'$ reduces to the Hermitian form given in Eq.~\eqref{eq:tensor_form} when the background magnetic field $\mathbf B_0$ vanishes, but also when $\mathbf B_0\parallel \boldsymbol \Omega$. As it happens that the latter corresponds precisely to the aligned rotator case previously studied for plasmas~\cite{gueroult2019,gueroult2020}, we focus on this configuration and find
\begin{equation} \label{eq:dielectric_constant}
    \chi_{\parallel}'(\omega') = -\frac{\omega_{pe}^2}{\omega'^2-\omega_0^2}
\end{equation}
the Lorentz's constant dielectric susceptibility, and
\begin{subequations} \label{eq:dielectric_tensor}  
\begin{align}
    {\chi'_\perp}(\omega') &= \frac{-(\omega'^2+\Omega^2-\omega_0^2)\omega_{pe}^2}{\left(\omega'^2+\Omega^2-\omega_0^2\right )^2 - \omega'^2 \left(2\Omega-\Omega_{ce}\right)^2}, \\
    {\chi'_\times}(\omega') &= \frac{\omega'(2\Omega-\Omega_{ce})\omega_{pe}^2}{\left(\omega'^2+\Omega^2-\omega_0^2\right )^2 - \omega'^2 \left(2\Omega-\Omega_{ce}\right)^2}.
\end{align}
\end{subequations}
In the limit of zero magnetic field ($\Omega_{ce}=0$) we recover the formulas found by Shiozawa in~\cite{shiozawa1973} (other than for the unexplained absence of the centrifugal contribution in the numerator of $\chi'_{\perp}$ but not elsewhere in Ref.~\cite{shiozawa1973}). If one further considers $\Omega\rightarrow0$, then $\omega'\rightarrow\omega$, and one verifies as expected that ${\chi'_\perp}(\omega')\rightarrow \chi_{\parallel}(\omega)$ and ${\chi'_\times}(\omega')\rightarrow 0$, that is the classical Lorentz oscillator isotropic response.

\subsection{\label{sub:fluid}Rotating magnetized plasma}

A plasma cannot be modeled as a superposition of Lorentz oscillators because electrons are free instead of bound to a nucleus, inducing collective effects that must be modelled through kinetic or fluid descriptions. Here we consider the simple case of a collisionless cold plasma (index $e$ and $i$ for electrons and ions, respectively). The momentum conservation equation for each fluid $s$  in its rest-frame in response to a perturbation ($\mathbf{E}', \mathbf{B}'$) then writes
\begin{equation}
    \frac{d\mathbf v'_s}{dt}  = \frac{q_s}{m_s}\left[\mathbf E'+\mathbf v'_s\times \tilde{\mathbf B}'\right]
\end{equation}
where $m_s$ and $q_s$ are the mass and charge of the particles of species $s$, and $\mathbf v_s'$ is the fluid's Eulerian velocity at a given fixed position $\mathbf r_s'$ in $\Sigma'$. Here the effect of rotation is hidden in the total derivative, which following Thyagaraja and McClements~\cite{thyagaraja2009} writes
\begin{equation}
    \frac{d\mathbf v'_s}{dt}  = \partial_{t} \mathbf v'_s + \mathbf v'_s\cdot\boldsymbol\nabla'\mathbf v'_s+2\boldsymbol\Omega\times\mathbf v'_s + \boldsymbol\Omega\times[\boldsymbol\Omega\times\mathbf r'_s].
\end{equation}
Here ${\partial_{t} \mathbf v'_s + \mathbf v'_s\cdot\boldsymbol\nabla'\mathbf v'_s}$ is the classical Lagrangian derivative (capturing fluid convection), and the next two terms are the Coriolis and centrifugal contributions. Similarly to what was done for the linearized perturbation obtained for dielectrics, we then invert this relation (see Appendix~\ref{app:inversion}) and sum the response of each species to get the plasma dielectric tensor
\begin{subequations} \label{eq:general_dielectric_tensor_plasma}
\begin{equation} 
    \underline{\boldsymbol\chi}'= \sum_s\left[\underline{\boldsymbol\chi}'_{\dagger s}-\frac{\underline{\boldsymbol\chi}'_{\dagger s}\cdot \boldsymbol\varpi_s\boldsymbol\varpi_s \cdot \underline{\boldsymbol\chi}'_{\dagger s}}{1+{}^\textsf{T}\boldsymbol\varpi_s\cdot\underline{\boldsymbol\chi}'_{\dagger s}\cdot\boldsymbol\varpi_s}\right]
\end{equation}
where 
\begin{equation}
    \underline{\boldsymbol\chi}'_{\dagger s} = -\omega_{ps}^2\frac{\omega'^2\underline{\mathbf I} - \boldsymbol\Omega_s^*\boldsymbol\Omega_s^*-i\omega' \boldsymbol\Omega^*\times\underline{\mathbf I}}{\omega'^2-{\Omega_s^*}^2},
\end{equation}
\begin{equation}
    \boldsymbol\varpi_s\boldsymbol\varpi_s = -\frac{1}{\omega_{ps}^2}\Big[\Omega^2\underline{\mathbf I} - \boldsymbol\Omega\boldsymbol\Omega\Big],
\end{equation}
\end{subequations}
and we have introduced ${\boldsymbol\Omega_s^* = 2\boldsymbol\Omega+q_s\mathbf B_0'/m_s}$ the effective cyclotron frequency generalized to any species. Taking again $\mathbf B_0\parallel\boldsymbol{\Omega}$ to obtain the simpler form Eq.~\eqref{eq:tensor_form}, we finally find
\begin{equation} \label{eq:cold_plasma_dielectric_constant}
    \chi_{\parallel}'(\omega') = - \sum_s\frac{\omega_{ps}^2}{\omega'^2} \sim  - \frac{\omega_{pe}^2}{\omega'^2}
\end{equation}
the static cold plasma's susceptibility constant and
\begin{subequations} \label{eq:cold_plasma_tensor}
\begin{align}
    \chi'_{\perp}(\omega') &= \sum_s\frac{-(\omega'^2+\Omega^2)\omega_{ps}^2}{(\omega'^2+\Omega^2)^2 - \omega'^2(2\Omega+\varepsilon_s\Omega_{cs})^2}, \\ 
    \chi'_{\times}(\omega') &= \sum_s\frac{\omega'(2\Omega+\varepsilon_s\Omega_{cs})\omega_{ps}^2}{(\omega'^2+\Omega^2)^2 - \omega'^2(2\Omega+\varepsilon_s\Omega_{cs})^2},
\end{align}
\end{subequations}
where ${\varepsilon_s=q_s/|q_s|}$, ${\omega_{ps} = [q_{s}^2N_s/(\epsilon_0m_s)]^{1/2}}$ is the plasma frequency and  $\Omega_{cs} = |q{_s}|B_0/m_s$ is the unsigned cyclotron frequency. We verify that neglecting the ionic contribution gives back the results of Shiozawa~\cite{shiozawa1974}, and that keeping only the contribution of the Coriolis force leads to the rest-frame susceptibilities already identified by Engels \& Verheest~\cite{engels1975}.

\subsection{\label{sub:analogy}Analogies}

Comparing Eqs.~\eqref{eq:dielectric_constant} and \eqref{eq:dielectric_tensor} with Eqs.~\eqref{eq:cold_plasma_dielectric_constant} and \eqref{eq:cold_plasma_tensor} shows that the rotating Lorentz oscillator and the rotating magnetized plasma models can be unified in a single model in the limit of high frequency wave (that is when the ionic contribution in plasmas can be neglected). Indeed, writing  
\begin{subequations} \label{eq:components_analogy}
\begin{align} 
    \chi_{\parallel}'(\omega') &= -\frac{\omega_{pe}^2}{\omega'^2-\omega_k^2}, \\
    {\chi'_\perp}(\omega') &= \frac{-(\omega'^2+\Omega^2-\omega_k^2)\omega_{pe}^2}{\left(\omega'^2+\Omega^2-\omega_k^2\right )^2 - \omega'^2 \left(2\Omega-\Omega_{ce}\right)^2}, \\
    {\chi'_\times}(\omega') &= \frac{\omega'(2\Omega-\Omega_{ce})\omega_{pe}^2}{\left(\omega'^2+\Omega^2-\omega_k^2\right )^2 - \omega'^2 \left(2\Omega-\Omega_{ce}\right)^2},
\end{align}
\end{subequations}
one recovers the rotating Lorentz oscillator susceptibility Eqs.~\eqref{eq:dielectric_constant} and \eqref{eq:dielectric_tensor} for $\omega_k=\omega_0$ and the rotating magnetized plasma susceptibility Eqs.~\eqref{eq:cold_plasma_dielectric_constant} and \eqref{eq:cold_plasma_tensor} for $\omega_k=0$. Although not surprising in that $\omega_k=0$ corresponds to harmonic oscillator with a zero frequency restoring force, that is a free particle, this notation will prove convenient to study the relevance of the Coriolis-Faraday equivalence in the next section.

\section{\label{sec:drag}Inertial corrections to polarization drag}

The macroscopic models capturing the effect of rotation derived in Sec.~\ref{sec:models} can now be plugged in the formulas for laboratory frame circular birefringence theory recalled in Sec.~\ref{sec:theory} to determine the refractive indices of \textsf{L} and \textsf{R} waves in the {aligned rotator} configuration (${\mathbf k \parallel \mathbf B_0 \parallel \boldsymbol\Omega}$) taking into account inertial effects, and that for any wave frequency. The impact of inertial effect on polarization drag is then quantified by introducing the inertial rotation angle $\delta_\textsf{iner}\equiv\delta_\textsf{mech}-\delta_\textsf{rig}$, where $\delta_\textsf{mech}$ and $\delta_\textsf{rig}$ are respectively the rotation polarization per unit length along $\mathbf{k}$ with and without fictitious forces. 

\subsection{High frequency Faraday rotation in a static magnetized medium} \label{sub:faraday}

Before studying the rotatory case, and because it will come in handy to discuss later the relevance of the Coriolis-Faraday analogy, let us first show how classical results on Faraday rotation in the presence of a background magnetic field (see \emph{e.g.} Sommerfeld~\cite{sommerfeld1964}) can be recovered from Eq.~(\ref{eq:components_analogy}). 

Assuming a non-rotating medium (${\Omega=0}$), the two frames of reference $\Sigma$ and $\Sigma'$ merge and we can drop temporarily the prime notation. Eq.~\eqref{eq:birefringence} then reduces to the well-known dispersion relation of electromagnetic waves in a gyrotropic medium ${n_\textsf{R/L}^2(\omega) = 1 + \chi_\perp(\omega) \pm \chi_\times(\omega)}$. Taylor expanding Eq.~(\ref{eq:components_analogy}) for $\Omega_{ce}/\omega\ll1$, it comes from Eq.~\eqref{eq:polarization_rotation_angle}
\begin{equation} \label{eq:theta_faraday}
     \delta_\textsf{mag}(\omega,B_0) = \frac{e^3}{2c\epsilon_0m_e^2} \frac{N_e B_0}{n_\phi}\frac{\omega^2}{(\omega_k^2-\omega^2)^2} + \mathcal{O}\left(\left[\frac{\Omega_{ce}}{\omega}\right]^2\right)
\end{equation}
with 
\begin{equation}
n_\phi = \sqrt{1+\chi_\parallel}=\left[1-\frac{\omega_{pe}^2}{\omega^2-\omega_k^2}\right]^{1/2}.
\end{equation}
This first order approximation of Faraday rotation, valid at high frequency, is commonly expressed by introducing the Verdet constant $V$ such that ${ \delta_\textsf{mag} = VB_0}$~\cite{verdet1854}. By identification one gets
\begin{equation} \label{eq:verdet_magnetized}
    V(\omega) = \frac{e}{2cm_e} \omega \frac{dn_\phi}{d\omega}.
\end{equation}
This relation was first derived by Becquerel~\cite{becquerel1897} and is also recovered by quantum models~\cite{berman2010}. 

\subsection{Validity of Coriolis-Faraday for high frequency waves in a non-magnetized rotating medium}

Our interest being in evaluating possible contributions of inertia to polarization drag, a natural first step is to examine how the predictions of the phenomenological models developed here compare with the original effort in this direction from Baranova \& Zel'dovich~\cite{baranova1979}. 

For reference Baranova \& Zel'dovich postulated that inertia should be the source of an extra contribution to polarization drag compared to Player's formula~\cite{player1976}, and surmised on the ground of Larmor's theorem that this contribution should be indiscernible - at least at high frequency - from the Faraday rotation computed for the equivalent magnetic field ${\mathbf B^*=-2m_e\boldsymbol\Omega/e}$. In other words and using our notation, Baranova and Zel'dovich predicted that
\begin{equation}
    \delta_\textsf{iner}(\omega,\Omega) = \delta_\textsf{mag}(\omega,-2m_e\Omega/e).
\end{equation}
Using this analogy, one could in principle determine the inertial contribution to polarization rotation from the definition ${ \delta_\textsf{mag} = VB_0}$ given the Verdet constant of the material. In fact, this is precisely what was done by Baranova \& Zel'dovich~\cite{baranova1979} to estimate inertia effects in Jones's experiment~\cite{jones1976}.

Consistent with Baranova \& Zel'dovich's hypothesis, we focus first on the electronic response only and consider the case of an unmagnetized medium. We thus take as our starting point Eq.~\eqref{eq:components_analogy} with $\Omega_{ce}=0$. Plugging these susceptibilities into Eq.~\eqref{eq:birefringence}, Eq.~\eqref{eq:polarization_rotation_angle} then yields to first order in $\Omega \ll \omega$ (\emph{i.~e} retaining only the Coriolis contribution)
\begin{subequations}\label{eq:theta_mech}
\begin{equation} 
    \delta_\textsf{mech} = {\frac{\Omega}c\left[n_{g} - \frac{1}{n_{\phi}} \right ]} + {\frac{\Omega(\xi-1)}c\left[n_{\phi} - \frac{1}{n_{\phi}} \right ]}
\end{equation}
where
\begin{equation}
    \xi(\omega) = \frac{\omega_k^2-2\omega^2}{\omega_k^2-\omega^2} \sim 
    \begin{cases} 
        1^- \text{ for } \omega_k \gg \omega \\
        2^+ \text{ for } \omega_k \ll \omega
    \end{cases}.
\end{equation}
\end{subequations}
Note though that, as pointed out earlier, this expansion is only valid if the refractive index ${n_\phi(\omega \mp\Omega)}$ does not vanishes when $\Omega/\omega$ goes to zero. 

Examining Eq.~(\ref{eq:theta_mech}), the first term on the right hand side corresponds precisely to the mechanical polarization rotation $\delta_\textsf{rig}$ derived by Player~\cite{player1976}. The second term on the right hand side in Eq.~(\ref{eq:theta_mech}) must thus correspond to the inertial (Coriolis as we limited ourselves to first order corrections in $\Omega$) contribution $\delta_\textsf{iner}$. Expanding this last term with the definition of $\xi$ and $n_{\phi}$ gives
\begin{equation}
    \delta_\textsf{iner}(\omega,\Omega) = - \frac{e^2}{c\epsilon_0 m_e}\frac{N_e\Omega}{n_\phi}\frac{\omega^2}{(\omega_k^2-\omega^2)^2},
\end{equation}    
which we verify from Eq.~\eqref{eq:theta_faraday} is precisely $\delta_\textsf{mag}(\omega,-2m_e\Omega/e)$. This is thus exactly the Faraday-Coriolis equivalence predicted by Baranova \& Zel'dovich~\cite{baranova1979}. 

Also, similarly to what was done through the definition of the Verdet constant in Eq.~\eqref{eq:verdet_magnetized}, the inertial correction can be rewritten as 
\begin{equation}
\delta_\textsf{iner}(\omega,\Omega) = -\frac{\Omega\omega}{c} \frac{dn_\phi}{d\omega}.
\end{equation}
One finds that this term term precisely balances the dispersive part of $\delta_\textsf{rig}$ obtained when expanding the group index ${n_{g} = n_\phi + \omega (dn_\phi/d\omega)}$ in Eq.~\eqref{eq:theta_mech}. The total specific polarization rotation $\delta_\textsf{mech}$ thus reduces to its kinematic part ${(\Omega/c)(n_\phi-n_\phi^{-1})}$, a result anticipated by Baranova \& Zel'dovich~\cite{baranova1979} and explained in terms of angular momentum exchange by Woerdman \emph{et al.}~\cite{woerdman1992}. 

In summary, our results are consistent with the Coriolis-Faraday term for high frequency polarization drag in an unmagnetized rotating medium, supporting Baranova \& Zel'dovich predictions in this regime~\cite{baranova1979}.

\subsection{Intrinsic limits of the Faraday-Coriolis equivalence: magnetized rotating medium}

Although attractive and effective in the high-frequency regime considered above, a closer examination reveals a number of intrinsic limitations to the Coriolis-Faraday analogy to quantify the effect of inertia on polarization drag. 

A first limitation of the Faraday-Coriolis analogy, or at least its implementation using the Verdet constant as done by Baranova \& Zel'dovich~\cite{baranova1979}, arises for wave frequencies such that the Faraday rotation due to the equivalent magnetic field ${\mathbf B^*=-2m_e\boldsymbol\Omega/e}$ is non-linear in $\mathbf B^*$. From Sec.~\ref{sub:faraday} this occurs when $e B^*/m_e$ approaches $\omega$, that is when $2\Omega$ becomes comparable to $\omega$. Yet, because for most materials and regimes we are typically interested the wave frequency is orders of magnitude larger than the rotation frequency, this limit is not particularly constraining. 

A second and more limiting problem is found when the properties of a medium at rest are such that two circularly polarised eigenmodes are found for a given wave frequency $\omega$, but that the properties of this same media in the presence of the equivalent magnetic field $\mathbf B^*$ do not allow for at least one of these modes at this same frequency. In this case one is indeed left with polarization drag but no corresponding equivalent Faraday rotation. This is for instance what is found in an underdense (\emph{i.e.} $\omega_{pe}\ll\Omega_{ce}$) rotating magnetized plasma in the limit that $\omega_{pe}\gg\Omega$. 

\begin{figure}
    \includegraphics[width=1.1\linewidth]{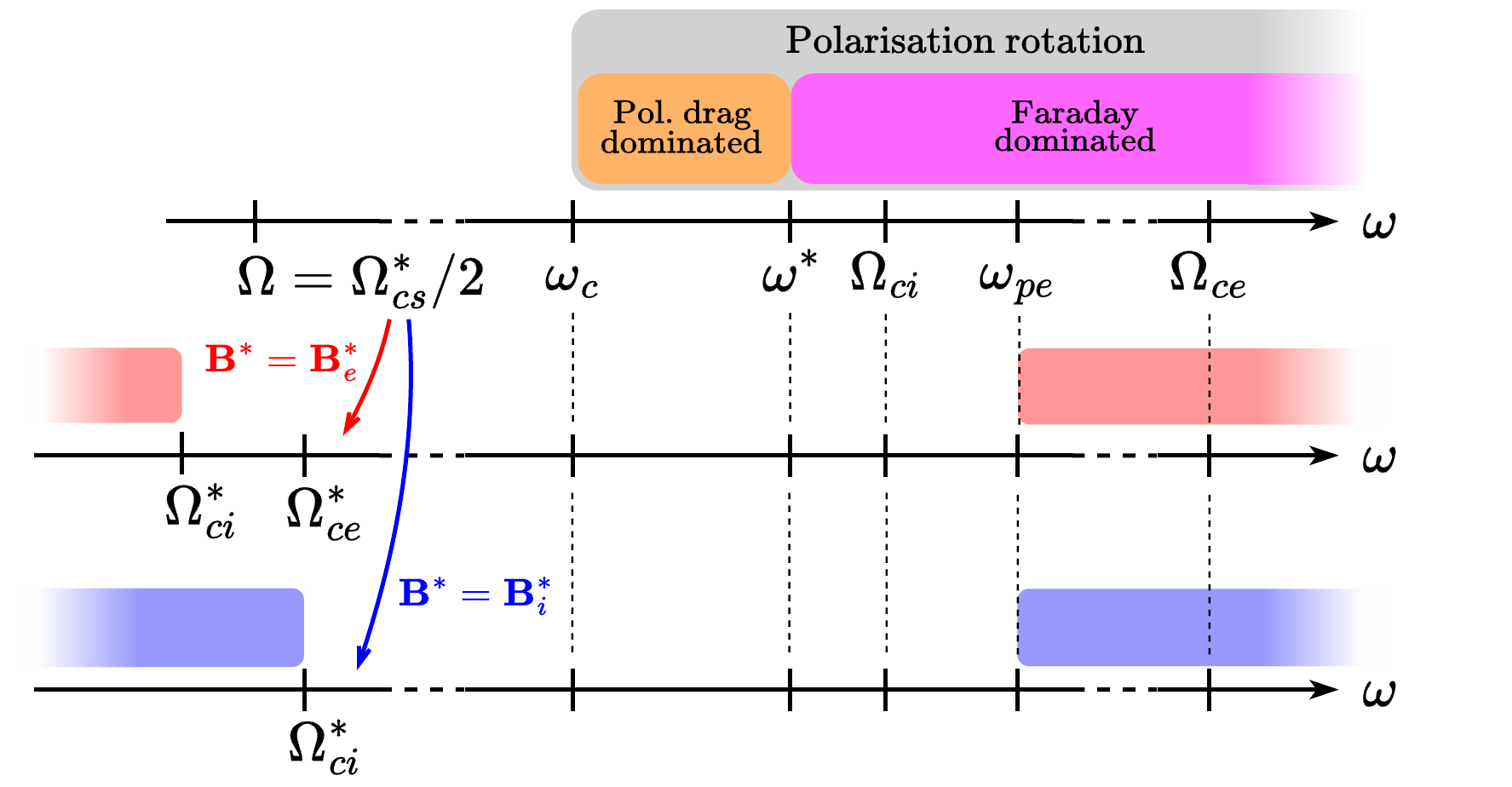}
    \caption{\label{fig:limit} Bandwidth where polarization rotation is at play in an underdense rotating magnetized plasma (grey, top), versus those for which the equivalent Faraday rotation would be defined considering the Larmor equivalence for electrons (red, middle) and ions (blue, bottom). For $\omega_c\leq\omega\leq\omega_{pe}$ polarization rotation in an underdense rotating magnetized plasma occurs, but the equivalent Faraday rotation is not defined.}
\end{figure}

To see this, recall that the cutoffs for right- and left-circularly polarised eigenmodes propagating along $\mathbf{B}_0$ in an electron-ion plasma write
\begin{subequations}
\begin{equation}
\omega_\textsf{R}  = \frac{\Omega_{ce}-\Omega_{ci}}{2}+\left[\left(\frac{\Omega_{ce}+\Omega_{ci}}{2}\right)^2+{\omega_{pe}}^2+{\omega_{pi}}^2\right]^{1/2}
\label{eq:rcutoff}
\end{equation} 
and 
\begin{equation}
\omega_\textsf{L}  = \frac{\Omega_{ci}-\Omega_{ce}}{2}+\left[\left(\frac{\Omega_{ce}+\Omega_{ci}}{2}\right)^2+{\omega_{pe}}^2+{\omega_{pi}}^2\right]^{1/2}.
\label{eq:lcutoff}
\end{equation}  
\end{subequations}
For the underdense plasma considered here ${\omega_\textsf{R}\rightarrow\Omega_{ce}}$ and ${\omega_\textsf{L}\rightarrow\Omega_{ci}}$, and as a result propagation of both modes is possible at any frequency in the limit $\omega_{pe}/\Omega_{ce}\rightarrow 0$. In Refs.~\cite{gueroult2019,gueroult2020} it was shown that this holds true when accounting for rotation, other than for the existence in this case of a mechanically induced low frequency cutoff \begin{equation}
\label{eq:cutoff_omega_c}
\omega_c\sim\left[\omega_{pe}^2\Omega\right]^{1/3}\ll\Omega_{ci}.
\end{equation}
As illustrated in Fig.~\ref{fig:limit}, it was further shown that polarization rotation occurs primarily due to the mechanical rotation just above the cutoff, and conversely primarily due to the magnetic field at higher frequency, with a crossover for a frequency \begin{equation}
\label{eq:crossover_omega_star}
\omega^{\star} \sim\eta\left[\Omega_{ce}^3\Omega\right]^{1/4}
\end{equation}
with $\eta^2$ the electron to ion mass ratio. Note that this was done neglecting inertia corrections, but this result will be confirmed \textit{a posteriori} in this study. Meanwhile, by the definition of ${\mathbf B^*=2m_{s}\boldsymbol\Omega/q_{s}}$, one finds $\Omega_{cs}^*=2q_{s}/|q_{s}|\Omega$, with $\Omega_{cs}^*$ the cyclotron frequency computed for the equivalent field $B^*$. If one further makes the reasonable assumption $\omega_{pe}\gg\Omega$, then necessarily $\omega_{pe}\gg\Omega_{cs}^*$, which implies that the plasma is overdense for the equivalent magnetic field $B^*$. In this case, using Eqs.~\eqref{eq:rcutoff} and \eqref{eq:lcutoff}, both eigenmodes are only found below $\omega_\textsf{L}\rightarrow \Omega_{ci}^*=2\Omega$ and above $\omega_\textsf{R}\rightarrow\omega_{pe}$. Accordingly the Faraday rotation due to the equivalent field $B^*$ is only defined for $\omega\leq 2\Omega$ and $\omega\geq\omega_{pe}$. Putting these pieces together, we are left with the finding that although polarization drag is at play in our underdense plasma for $\omega_c\leq\omega\leq\Omega_{ci}$ (top panel in Fig.~\ref{fig:limit}), the Coriolis-Faraday equivalence cannot be used to quantify inertial corrections in this frequency band. Because this is precisely in this frequency band that both polarization drag in the rotating magnetosphere of pulsars~\cite{gueroult2019} and enhanced mechanical polarization drag~\cite{gueroult2020} are expected, we will now examine mechanical corrections using the models developed in Sec.~\ref{sec:models}.

\subsection{Inertial corrections to low frequency polarization drag in a rotating magnetized plasma}

Having seen that the Faraday-Coriolis can not account for inertia corrections to low frequency polarization drag in a rotating magnetized plasma, we now would like to assess whether inertia corrections are important in this regime. For this we take as a reference the underdense plasma properties already considered in Ref.~\cite{gueroult2020}, which are recalled in Table~\ref{tab:baseline}.

\begin{table}
    \caption{\label{tab:baseline}%
    Baseline plasma parameters used to estimate inertial corrections to polarization drag in an underdense plasma, taken from Ref.~\cite{gueroult2020}. }
    \begin{ruledtabular}
        \begin{tabular}{lll}
            Plasma density & $N_e$ & $10^{20}$ m$^{-3}$ \\
            Magnetic field & $B_0$ & $10^{3}$ T \\
            Rotation frequency & $\Omega$ & $100$~Hz\\
        \end{tabular}
    \end{ruledtabular}
\end{table}

\begin{figure}
    \includegraphics[width=0.8\linewidth]{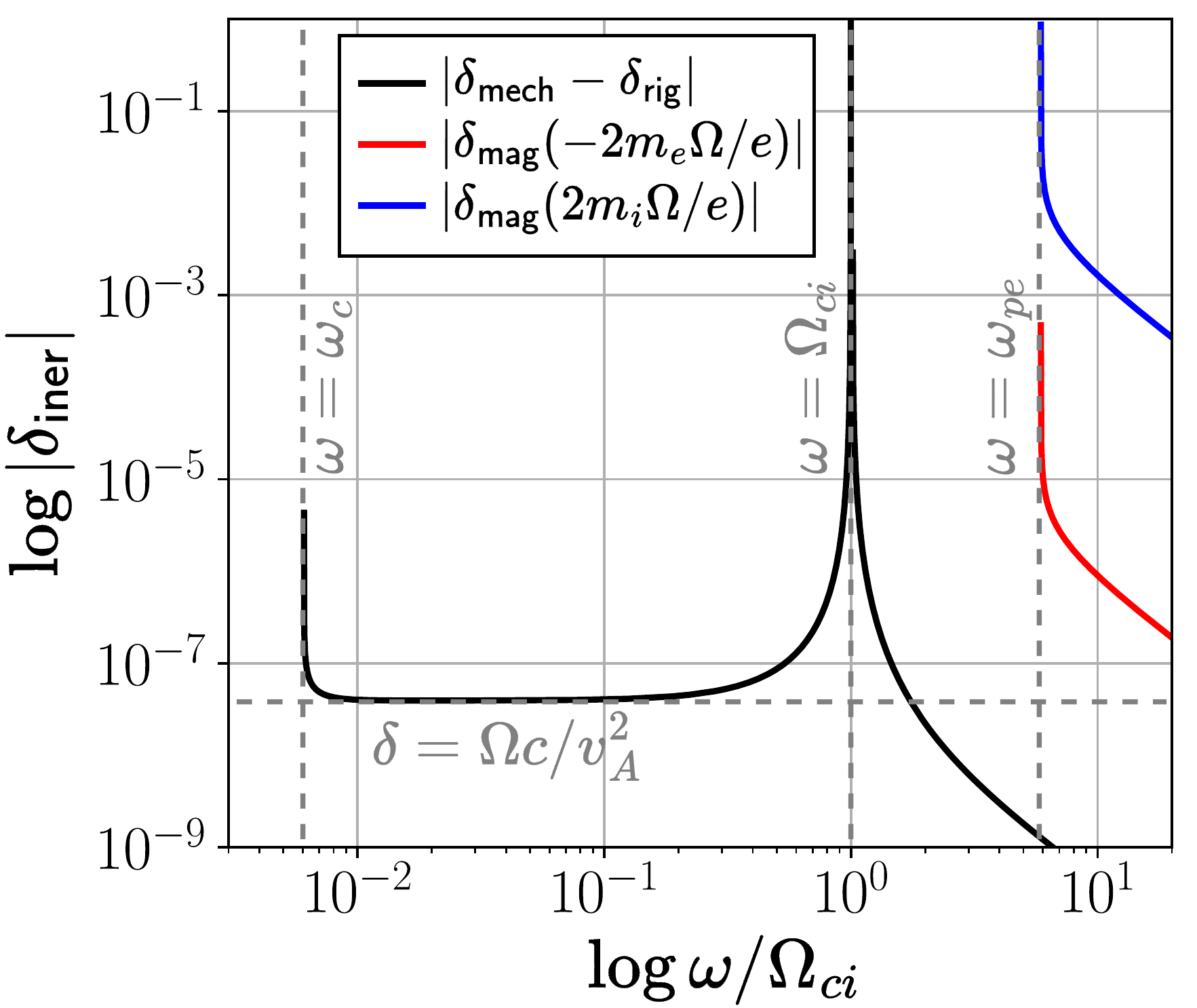}
    \caption{\label{fig:faraday_coriolis} Inertial correction to polarization drag computed from Eq.~\eqref{eq:index_magnetized_plasma} for the plasma parameters given in Table~\ref{tab:baseline} (in black). The horizontal grey dashed curve corresponds to the asymptotic result $- \Omega c/v_A^2$.  The red and blue curves represent the correction deduced from the Faraday-Coriolis equivalence for these same parameters and considering the equivalent magnetic field $\mathbf{B}^*$ for respectively electrons and ions, confirming the non applicability of this analogy for low frequency polarization drag. }
\end{figure}

Plugging the dielectric tensor components of the magnetized plasma model Eqs.~\eqref{eq:cold_plasma_dielectric_constant}-\eqref{eq:cold_plasma_tensor} from Sec.~\ref{sub:fluid} in Eq.~\eqref{eq:birefringence} yields the wave indices for \textsf{L}- and \textsf{R}- circularly polarized modes 
\begin{equation} \label{eq:index_magnetized_plasma}
    n_\textsf{R/L}^2(\omega) = {1  - \sum_s\frac{\omega_{ps}^2}{\omega'^2} \left[\frac{\omega'^3}{\omega(\omega^2\pm\varepsilon_s\Omega_{cs}\omega')} \mp \frac{\Omega}{\omega} \right ]}
\end{equation}
with again $\omega'=\omega \mp \Omega$ the rotating Doppler shift. For frequency below the ion cyclotron frequency but above the mechanically induced cutoff $\omega_c$, \emph{i.e.} $\omega_c \ll \omega \ll \Omega_{ci}$, one shows that this lead to a specific rotatory power to first order in $\Omega$
\begin{equation}
    \delta_\textsf{mech} \sim \delta_\textsf{rig} - \frac{\Omega c}{v_A^2}.
\end{equation}
In other words the inertial correction to polarization drag writes  $\delta_\textsf{iner} = - \Omega c/v_A^2$ with $v_A$ the Alfv{\'e}n speed. Note that it is independent of the wave frequency to this order. This behaviour and the scaling are confirmed when obtaining numerically the polarization rotation angle obtained from Eq.~\eqref{eq:index_magnetized_plasma}, as shown in Fig.~\ref{fig:faraday_coriolis}. We also verify in this same figure that the inertial correction cannot be deduced from the Faraday-Coriolis equivalence, whether it is considering ions or electrons. Indeed, as previously discussed in either case the Faraday-Coriolis term in not defined below $\omega_{pe}$.

This inertial contribution should now be compared to the two scalings valid below the ion-cyclotron frequency~\cite{gueroult2020}, namely $\delta_\textsf{rig}\sim -\omega_{pe}^2\Omega/(2c\omega^2)$ for the mechanically dominated polarization rotation at low-frequency, and $\delta_\textsf{rig}\sim \omega_{pi}^2\omega^2/(2c\Omega_{ci}^3)$ for the higher frequency Faraday dominated polarization rotation. Inertial corrections to the mechanically dominated region thus write
\begin{equation}
     \delta_\textsf{iner} = \frac{2\eta^2\omega^2}{\Omega_{ci}^2}\delta_\textsf{rig}.
     \label{eq:iner_scaling_LF}
\end{equation}
This implies that inertial corrections add up to the rigid solution. Since this is valid for low-frequency $\omega\ll\Omega_{ci}$, these corrections are however small, and in effect independent of $\Omega$ when neglecting second order effect (centrifuge terms). Meanwhile, inertial corrections to the Faraday dominated polarization rotation write
\begin{equation}
     \delta_\textsf{iner} = -\frac{2\Omega_{ci}\Omega}{\omega^2}\delta_\textsf{rig}.
     \label{eq:iner_scaling_HF}
\end{equation}
Inertial corrections thus subtracts from the rigid solution in this limit, with inertial corrections in the same direction as the mechanical contribution but opposite to Faraday rotation. We note that inertial corrections can be here more marked, especially for larger $\Omega$ approaching $\Omega_{ci}$.  

\begin{figure}
    \includegraphics[width=1\linewidth]{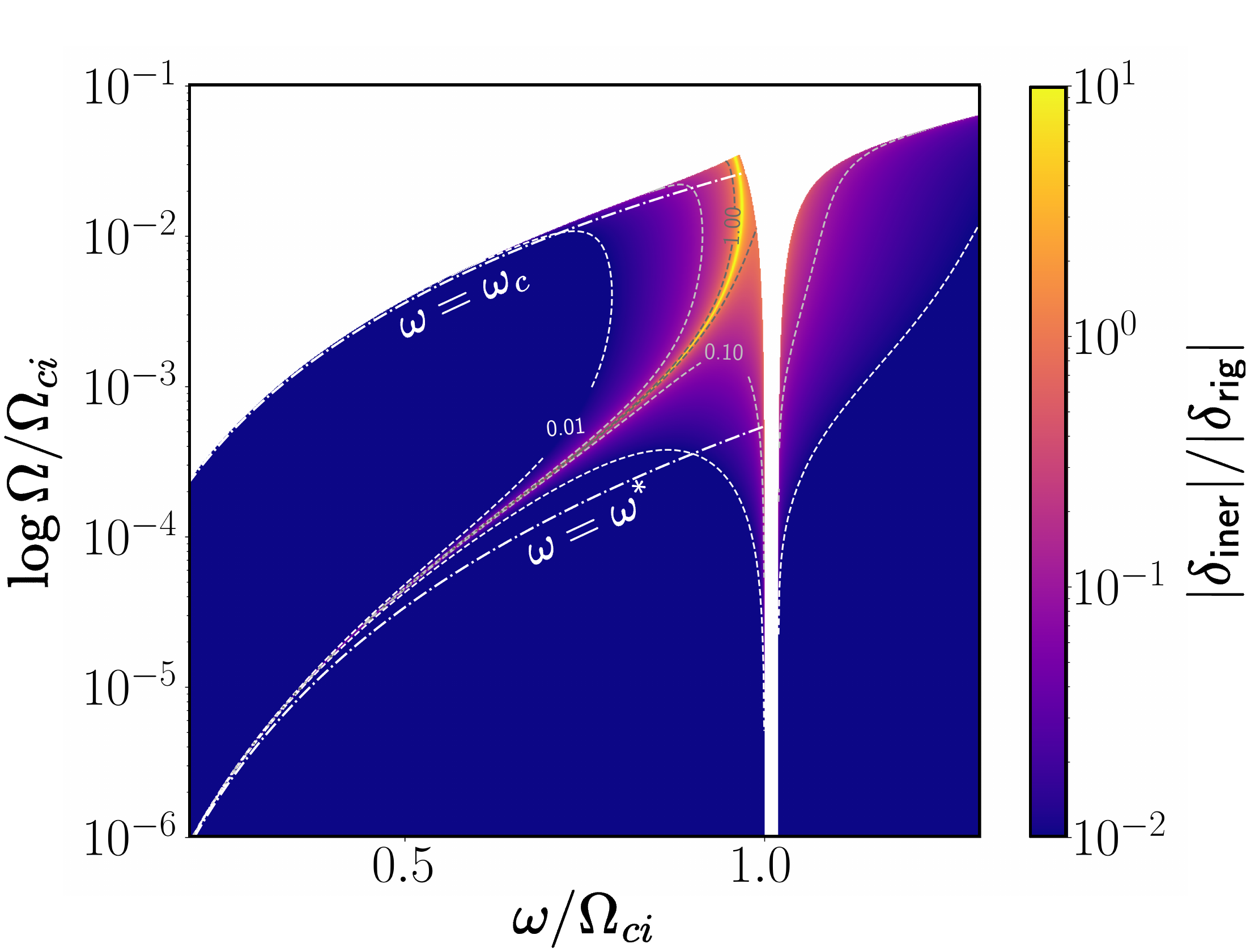}
    \caption{\label{fig:inertial_effects} Contribution of inertial effects to the polarization rotation relative to the rigid body polarization rotation (\textit{i.e.} without inertial effects), for an angular velocity $\Omega$ between $10^{-4}\Omega_{ci}$ and $\Omega_{ci}/4$. Plasma conditions ($N_e,B_0$) correspond to those given in Table~\ref{tab:baseline}. The white dash-dotted lines represent the cutoff $\omega_c$ and crossover $\omega^{\star}$ frequency without inertial corrections, as given in Eqs.~\eqref{eq:cutoff_omega_c} and \eqref{eq:crossover_omega_star}. The large values near $\omega^{\star}$ are simply the consequence that $\delta_\textsf{rig}=0$ at the crossover frequency. }
\end{figure}

These trends are finality confirmed when solving numerically for the relative contribution of inertial corrections to polarization rotation as a function of the angular frequency $\Omega$ as shown in Fig.~\ref{fig:inertial_effects}. Consistent with Eqs.~\eqref{eq:iner_scaling_LF} and \eqref{eq:iner_scaling_HF}, one verifies that the ratio $\delta_\textsf{iner}/\delta_\textsf{rig}$ remains independent of $\Omega$ just above the cutoff $\omega_c$, that is in the polarization drag dominated regime, but grows with $\Omega$ above the crossover frequency $\omega^{\star}$. Furthermore, for the latter and for $\Omega$ approaching $\Omega_{ci}$ corrections of $\mathcal{O}(1)$ are observed as expected. Note though that because as shown in Eqs.~\eqref{eq:cutoff_omega_c} and \eqref{eq:crossover_omega_star} both $\omega_c$ and $\omega^{\star}$ grow with $\Omega$, the frequency band over which polarization rotation is defined and the region over which polarization drag dominates over Faraday rotation varies with $\Omega$. In fact, this dependence on $\Omega$ is actually more complex as these two frequencies are also affected by inertia. Yet, while the crossover frequency $\omega^{\star}$ clearly deviates from the simple scaling \eqref{eq:crossover_omega_star} for large rotation, and even starts to decrease with $\Omega$ past a point, we observe in Fig.~\ref{fig:inertial_effects} that inertial corrections to the cutoff frequency $\omega_c$ are minimal.

To sum up, because both circularly polarized modes propagate below $\Omega_{ci}$ along the axis of an underdense rotating magnetized plasma in an aligned rotator configuration ($\mathbf{k}||\boldsymbol\Omega||\mathbf{B}_0$), polarization rotation occurs and could in principle be affected by inertia. At such frequency though, the Faraday-Coriolis equivalence is not applicable as the frequency falls within a cutoff band of the circularly polarized modes obtained for the equivalent magnetic field $\mathbf{B}^\star$. Computing this inertial contribution from the phenomenological models developed in Sec.~\ref{sec:models}, we show however that corrections remain small in the polarization drag dominated regime no matter how large the rotation, and that corrections are also small in the Faraday dominated regime other than when $\Omega/\Omega_{ci}=\mathcal{O}(1)$. These results suggest that inertial corrections can be safely neglected when considering the low-frequency enhancement of polarization rotation near the mechanically induced cutoff $\omega_c$~\cite{gueroult2020}.

\section{\label{sec:conclusions}Conclusions}

In this study we examined possible inertial corrections to the polarization drag - or mechanically-induced birefringence - experienced by light propagating along the axis of rotation of a medium. More specifically, we considered the effect associated with the inertial corrections induced by rotation to the dielectric response of a rotating medium in its rest-frame. To do so we included the fictitious Coriolis and centrifugal forces into classical microscopic models for the dynamics of a rotating Lorentz dielectric and of a cold rotating plasma. The associated corrections to the circular birefringence as observed in the laboratory frame where then deduced based on existing theory~\cite{player1975,gueroult2019} relating rest-frame dielectric properties to polarization drag in the laboratory frame.

These new models confirm that for high frequency waves the inertial correction is precisely the one obtained if considering instead of rotation the Faraday rotation associated with the magnetic field obtained from Larmor's theorem, as originally suggested by Baranova \& Zel'dovich~\cite{baranova1979}. Yet, these same models show not only that this simple analogy fails at low frequency (as already anticipated by Baranova \& Zel'dovich), but also in rotating magnetized media such as a rotating magnetized plasma. In this particular case it is however demonstrated that inertial corrections to polarization drag below the ion cyclotron frequency are in general negligible, therefore confirming \emph{a posteriori} the results obtained previously when neglecting these effects~\cite{gueroult2020}. 

Finally, although this work does not provide new elements that could immediately weigh in on the apparent contradiction between Jones's experimental observations~\cite{jones1976} and Baranova \& Zel'dovich's predictions~\cite{baranova1979} for a rotating glass rod as one would need for that microscopic models for a rotating molecular dielectric, we note that the results obtained here could in principle be used to confirm experimentally the existence of inertial corrections to polarization drag in the case of a rotating plasma.

\begin{acknowledgments}
    This work was supported by the French Agence Nationale de la Recherche (ANR), under grant ANR-21-CE30-0002 (project WaRP). JL acknowledges the support of ENS Paris-Saclay through its Doctoral Grant Program.
\end{acknowledgments}

\appendix

\section{Dyadic notation and identities}
\label{app:dyadic}

Let $\mathbf a$ and $\mathbf b$ be two vectors of $\mathbb R^3$. We note $\mathbf a \mathbf b$ the dyadic product generating the second order tensor $(ab)_{ij}=a_ib_j$. The dyadic unit $\underline{\mathbf I}$, whose corresponding matrix is the unit matrix, verifies ${\mathbf a \cdot \underline{\mathbf I} = \underline{\mathbf I} \cdot \mathbf a = \mathbf a}$. Now let $\underline{\mathbf F}$ be a dyad and $\mathbf F_j$ its $j$-th column written as a vector, then the $j$-th column of the vector-dyad cross product $\mathbf a \times \underline{\mathbf F}$ is given by
\begin{equation}
    (\mathbf a \times \underline{\mathbf F})_j= \mathbf a \times \mathbf F_j.
\end{equation}
With these definitions and with $\mathbf c$ a vector of $\mathbb R^3$
\begin{align}
    \mathbf a \times (\mathbf b \times \mathbf c) &= [\mathbf a \times (\mathbf b \times \underline{\mathbf I})] \cdot \mathbf c \label{eq:dyad1} \\
    \mathbf a \times (\mathbf b \times \underline{\mathbf I}) &= \mathbf b \mathbf a - (\mathbf a \cdot \mathbf b) \underline{\mathbf I} \label{eq:dyad2}
\end{align} 
Using Eqs.~\eqref{eq:dyad1} and~\eqref{eq:dyad2} with ${\mathbf a = \mathbf b = \boldsymbol{\Omega}}$ and ${\mathbf c = \boldsymbol r_e'}$, the centrifugal contribution may be rewritten as
\begin{equation*}
    \boldsymbol\Omega\times(\boldsymbol\Omega\times\boldsymbol r'_e) = (\boldsymbol\Omega\boldsymbol\Omega - \Omega^2 \underline{\mathbf I})\cdot\boldsymbol{r}_e'.
\end{equation*}

\section{Inversion of the dielectric tensor}
\label{app:inversion}

Let us consider a tensor $\underline{\boldsymbol{\chi}}'$ whose inverse is
\begin{equation*}
    {\underline{\boldsymbol{\chi}}'}^{-1} = -\frac{1}{\omega_{pe}^2}\Big[(\omega'^2+\Omega^2-\omega_0^2){\mathbf I} + i\omega'\boldsymbol\Omega^*\times{\mathbf I}-\boldsymbol\Omega\boldsymbol\Omega\Big].
\end{equation*}
To get the general expression of $\underline{\boldsymbol{\chi}}'$, we invert first the Lorentz-Coriolis contributions for which a method is known (see \textit{e.g.}~\cite{engels1975}). Assuming the reduced tensor can be written as ${{\underline{\boldsymbol{\chi}}_\dagger}'=\alpha\underline{\mathbf I} + \beta\boldsymbol\Omega^*\boldsymbol\Omega^*+\gamma\boldsymbol\Omega^*\times\underline{\mathbf I}}$ with $\alpha,\beta,\gamma\in\mathbb R$, it comes from ${{\underline{\boldsymbol{\chi}}_\dagger}'\cdot {{\underline{\boldsymbol{\chi}}_\dagger}'}^{-1} = \underline{\mathbf I}}$ that 
\begin{equation*}
    \underline{\boldsymbol\chi}'_\dagger = -\omega_{pe}^2\frac{\omega'^2\underline{\mathbf I} - \boldsymbol\Omega^*\boldsymbol\Omega^*-i\omega' \boldsymbol\Omega^*\times\underline{\mathbf I}}{\omega'^2-{\Omega^*}^2}. 
\end{equation*}
Then, we use the Sherman–Morrison formula which states that for any second order tensor $\underline{\mathbf F}$ and $\mathbf a$ and $\mathbf b$ two vectors of $\mathbb R^3$, if $1 + {}^\textsf{T}\mathbf b\cdot\underline{\mathbf F}^{-1} \cdot \mathbf a \neq 1$ then
\begin{equation*}
    \left(\underline{\mathbf F} + \mathbf a \mathbf b\right)^{-1} = \underline{\mathbf F}^{-1} - \frac{\underline{\mathbf F}^{-1}\cdot\mathbf a\mathbf b \cdot \underline{\mathbf F}^{-1}}{1 + {}^\textsf{T}\mathbf b\cdot\underline{\mathbf F}^{-1} \cdot \mathbf a}.
\end{equation*}
Applying this relation with $\underline{\mathbf F} = \underline{\boldsymbol\chi}'_\dagger$ and $\mathbf a = \mathbf b = \boldsymbol{\varpi}$ gives Eqs.~\eqref{eq:general_dielectric_tensor} and \eqref{eq:general_dielectric_tensor_plasma}.

\end{document}